\documentclass[%
 reprint,
 amsmath,amssymb,
 aps,
]{revtex4-2}

\usepackage{graphicx}
\usepackage{dcolumn}
\usepackage{bm}
\usepackage{hyperref}
\usepackage{cleveref}
\usepackage{comment}
\usepackage{amsmath,color}
\usepackage{mathtools}
\usepackage{caption}
\usepackage{subcaption}
\usepackage{graphicx}
\usepackage{xcolor}

\begin{document}

\preprint{APS/123-QED}
\title{Paradoxical increase of capacity due to spurious overlaps in attractor networks}

\author{Marco Benedetti$^1$, Nicolas Brunel$^1$, Enzo Marinari$^2$ and Ulises Pereira-Obilinovic$^{3,*}$}
\affiliation{$\;^1$Department of Computing Sciences, Università Bocconi, Milano, Italy \\ $\;^2$Department of Physics, Università La Sapienza, Roma, Italy \\ $\;^3$Allen Institute for Neural Dynamics, Seattle, USA \\ $\;^*$alphabetical order.}
             
\begin{abstract}
In Hopfield-type associative memory models, memories are stored in the connectivity matrix and can be retrieved subsequently thanks to the collective dynamics of the network. In these models, the retrieval of a particular memory can be hampered by overlaps between the network state and other memories, termed `spurious overlaps’ since these overlaps collectively introduce noise in the retrieval process. In classic models, spurious overlaps increase the variance of synaptic inputs but do not affect the mean. We show here that in models equipped with a learning rule inferred from neurobiological data, spurious overlaps collectively reduce the mean synaptic inputs to neurons, and that this mean reduction causes in turn an increase in storage capacity through a sparsening of network activity. Our paper demonstrates a link between a specific feature of experimentally inferred plasticity rules and network storage capacity.
\end{abstract}

\maketitle

\section{\label{sec:level1}Introduction}

Attractor neural networks of the Hopfield type are a popular theoretical framework for learning and memory. In these models, memories are stored as fixed-point attractors of the network thanks to synaptic plasticity. Memories can be retrieved by providing an external input that leads the network in the basin of attraction of the corresponding fixed point attractor. This framework naturally explains the experimentally observed phenomenon of persistent activity in animals performing delayed-response tasks~\citep{fuster1971neuron,miyashita1988neuronal,funahashi1989mnemonic,goldman1995cellular,inagaki2019discrete,constantinidis2018persistent,khona2022attractor}.

In classical attractor networks such as the Hopfield model \cite{amit1985storing,amit1985storing}, retrieval of a particular memory is promoted by the overlap between the network state and this specific memory, but overlaps with other memories act as noise during the retrieval process, limiting the storage capacity of the model. When analyzing networks whose learning rules and neural nonlinearities are inferred from \emph{in vivo} data~\cite{lim2015inferring,pereira2018attractor}, we find however a qualitatively different behavior. Although overlaps with non-retrieved memories (termed \textit{spurious overlaps}) are individually small, their collective effect contributes an order-one reduction to the mean input current. This global suppression promotes sparser retrieval states, in which only a small fraction of neurons are strongly active. The resulting sparsity, in turn, sharply enhances the storage capacity of the network.

We develop a Dynamical Mean Field Theory~\cite{tirozzi1991chaos,pereira2023forgetting} that captures the influence of spurious overlaps on network dynamics. Using this theory, we derive analytical expressions characterizing retrieval properties and show that spurious overlaps can potentially increase memory capacity by reducing mean input currents to neurons. We show that this reduction occurs when the function that describes the dependence of the learning on post-synaptic firing rate (thereafter referred to as `POST function') has a negative average over stored patterns. In classical networks using covariance rules~\cite{sejnowski1977storing} such as the Hopfield model~\cite{hopfield1982neural}, this average is zero, so the effect of spurious overlaps on the mean input current vanishes. However, learning rules inferred from data have a significantly negative average of the POST function \cite{lim2015inferring,pereira2018attractor}, leading to a strong increase in memory capacity.
\section{The model}
We consider a network of $N$ analog neurons (e.g.~\cite{hopfield84,dayan2005theoretical}) whose rates evolve according to
\begin{equation}
        \dot{r}_i=-r_i+\Phi\Big(I_i+\sum_j c_{ij}J_{ij}r_j\Big).
        \label{eq:rate_dyn}
\end{equation}
In \cref{eq:rate_dyn}, $c_{ij}$ is a random Erdos-Renyi matrix, whose statistics follows
\begin{equation}
   P(c_{ij}=x)=c\delta(1-x)+ (1-c)\delta(x).
\end{equation} 

We concentrate here on the highly diluted regime $c\ll 1$. A sparse, asymmetric connectivity structure implies that correlations among firing rates of different neurons can be neglected at leading order, allowing a Mean Field Theory analysis. This regime is relevant for cortical microcircuits, where each neuron receives a large number of inputs ($K \sim 10^3$), but connection probabilities are low, on the order of $\sim$10\% in cortex~\cite{mason1991synaptic,markram1997physiology,holmgren2003pyramidal,thomson2007functional,lefort2009excitatory,campagnola2022local} and $\sim$1\% in hippocampus~\cite{guzman2016synaptic}.

The coupling matrix $J_{ij}$ is shaped by presenting to the network an infinite sequence of independently and identically distributed (i.i.d.) Gaussian input patterns $\xi^\mu_i \sim \mathcal{N}(0,1)$. This choice is made for consistency with \cite{lim2015inferring}, where the learning rule was inferred from \textit{in vivo} data. We assume that, during learning, inputs induce firing rates $\tilde{r}_i^\mu=\Phi(\xi^\mu_i)$, which in turn modify the coupling matrix according to
\begin{equation}
    J_{ij}^\mu=\Big(1-\frac{a}{Nc}\Big) J_{ij}^{\mu-1}+ \frac{b}{Nc}f(\tilde{r}_i^\mu)g(\tilde{r}_j^\mu),
\end{equation}
where $a$ and $b$ are parameters of the model that describe synaptic decay, and the strength of imprinting of new memories, respectively \cite{mezard1986solvable,pereira2023forgetting}. The transfer function $\Phi(\cdot)$ and the dependence of the learning rule on pre- ($g(\cdot)$, the PRE function) and post- ($f(\cdot)$, the POST function) synaptic firing rates are described by sigmoid functions $\Phi(x) =r_m/(1+\exp [-\beta_\Phi(x-h_0)]$, $f(x)=[2q_f-1+\tanh(\beta_f(x-x_f))]/2$, $g(x)=[2q_g-1+\tanh(\beta_g(x-x_g)) ]/2$. Sigmoidal $\Phi(x)$ and $f(x)$ have been shown to provide good fits to experimental data, using $r_m=76.2$ Hz, $h_0=2.46$, $\beta_{\Phi}=0.82$, $\beta_f=0.28$s, $x_f=26.6$Hz, and $q_f=0.83$. Importantly, these parameters lead to a negative average of the POST function,
\begin{equation}
\int P(\xi)f\big(\Phi(\xi)\big)d\xi<0,
\end{equation}
which ensures that familiar stimuli lead to lower average rates than novel ones~\cite{woloszyn2012effects,lim2015inferring}. For the function $g(x)$, consistent with \cite{pereira2018attractor}, we choose
$\beta_g=\beta_f$s,   $x_g=x_f$, and $q_g$ is chosen so that the PRE function has a zero average over the distribution of firing rates, i.e.~$\int P(\xi)g\big(\Phi(\xi)\big)d\xi=0$, leading to $q_g=0.89$. In the following, unless otherwise specified, we use the parameters that best fit the data, except $x_f=x_g=22$Hz.

The resulting palimpsestic coupling matrix is:
\begin{equation}
    J_{ij}=\frac{b}{Nc}\sum_{\mu=0}^\infty e^{-\frac{a \mu}{Nc}} f(\tilde{r}_i^\mu)g(\tilde{r}_j^\mu),
    \label{eq:coupling_matrix}
\end{equation}
with forgetting timescale $\tau=1/a$ and learning rate $b$. 
In Eq.~(\ref{eq:coupling_matrix}), the sum over $\mu$ is a sum over memories with different ages, $\mu=0$ being the most recent one, while $\mu \gg cN$ labels old memories that have been forgotten, due to the exponential decay term. We anticipate that the maximal age at which patterns can be retrieved scales linearly with $cN$ \cite{mezard1986solvable,derrida1987learning,pereira2023forgetting}, and define a rescaled age $s$ as $s=\mu/(cN)$.

\section{Mean Field Theory analysis}
Once the coupling matrix is assembled, one of the learned patterns is presented to the network in the form of an external field $I_i=\xi^\mu_i$, and the system is allowed to reach a steady state of the dynamics. Then, the external input is removed, setting $I_i=0$, and the dynamics proceeds until the system reaches a steady state. The characteristics of this latter “retrieval state” can be understood by studying the statistics of the local fields that drive the dynamics
\begin{equation}
    \lambda_i(t):=\sum_j c_{ij}J_{ij}r_j(t).
\end{equation}
As detailed in \cref{sec:Static_Mean_Field_Theory,sec:Dynamic_Mean_Field_Theory}, in the large $N$ limit fields $\lambda_i(t)$ behave as independent random variables 
\begin{equation}
    \lambda_i(t)\sim S+h_i(t) +e^{-\frac{a\mu}{Nc}}bf( \tilde{r}_i^\mu)m^\mu.
\end{equation}
where:

\begin{itemize}
    \item $m^\mu:=1/N\sum_i g( \tilde{r}_i^\mu) r_i$
    is an order parameter that quantifies the degree of overlap between the network state and stored pattern $\mu$. 
    \item $S:=\sum_{\alpha\neq\mu} e^{-\frac{a\alpha}{Nc}}b \langle f(\tilde{r})\rangle m^\alpha$ describes the cumulative effect on the mean input current of the overlaps between the network state and all other stored patterns $\xi^{\alpha\neq\mu}$. Here $\langle\cdot\rangle$ indicates averaging with respect to the Gaussian statistics of the stored patterns $\tilde{r}$.
    \item $h_i(t)$ is a Gaussian noise field, with correlation $\langle h_i(t)h_j(t)\rangle=\Delta_0 \delta_{ij}$ and $\lim_{\tau\rightarrow\infty}\langle h_i(t)h_j(t+\tau)\rangle =\Delta_{\infty}$. The order parameter $\Delta_0$ describes the variance of the field driving the network away from the memorized patterns, while $\Delta_{\infty}$ describes the correlation between the effective Gaussian noise field acting on a site at a given time, and the same field at much later times. If the steady state reached by the network is not a fixed point, but a chaotic attractor, we have $\Delta_0>\Delta_{\infty}$.
\end{itemize}
The order parameters $m^\mu,\,S,\,\Delta_0$ and $\Delta_\infty$ obey a set of self-consistency equations (see \cref{sec:Dynamic_Mean_Field_Theory}). Once they are solved, the average value of the spurious overlap with a memory of age $\alpha$ is given by
\small
\begin{equation}
    m^\alpha=\frac{b^2\,e^{-2\frac{a\alpha}{Nc}}}{2Nc}\frac{\big\langle\Phi''(\lambda)\big\rangle \big\langle g(\tilde{r}^\alpha) f^2(\tilde{r}^\alpha) \big\rangle \big\langle g^2(\tilde{r}^\alpha)\Phi^2(\lambda) \big\rangle}{1- b e^{-\frac{a\alpha}{Nc} }  \big\langle \Phi'(\lambda) \big\rangle \big\langle g(\tilde{r}^\alpha) f(\tilde{r}^\alpha) \big\rangle} \label{eq:spurious_overlaps_in_main}
\end{equation}
where $\langle\cdot\rangle$ indicates averaging with respect to the Gaussian statistics of $h$ and the stored patterns, and $\Phi', \Phi''$ are first and second the derivatives of $\Phi$. In the remainder of the text, index $\mu$ refers to the pattern the network is trying to retrieve, while $\alpha$, $\alpha\neq\mu,$ refers to all spurious overlaps.

 From the definition of $S$, one sees that learning rules with a zero average POST function, $\langle f(\tilde{r})\rangle=0$ give $S=0$, and thus no effect of spurious overlaps on the mean input currents (though they still contribute to the variance of these currents). Similarly, \cref{eq:spurious_overlaps_in_main} implies that if $\langle\Phi''(\lambda)\big\rangle=0$, spurious overlaps have zero average value, again leading to $S=0$. Classical attractor models~\cite{amit1985storing} typically satisfy these conditions; consequently, spurious overlaps make no average contribution \((S = 0)\).

On the other hand, learning rules inferred from \emph{in vivo} data~\cite{lim2015inferring} satisfy $\langle f(\tilde{r})\rangle < 0$, which leads to a decrease in average visual responses with familiarity. In addition, in realistic networks  $\langle\Phi''(\lambda)\big\rangle>0$, i.e. the transfer function operates typically in the supralinear regime. Hence, the $O(Nc)$ spurious overlaps $m^\alpha$, which individually have an average $O(1/Nc)$, lead to a finite average current contribution $S$ to the local fields, even though they individually vanish in the limit $N\to\infty$. 


\section{Beneficial effects of spurious overlaps}
\begin{figure*}[ht!]
\includegraphics[width=0.9\linewidth]{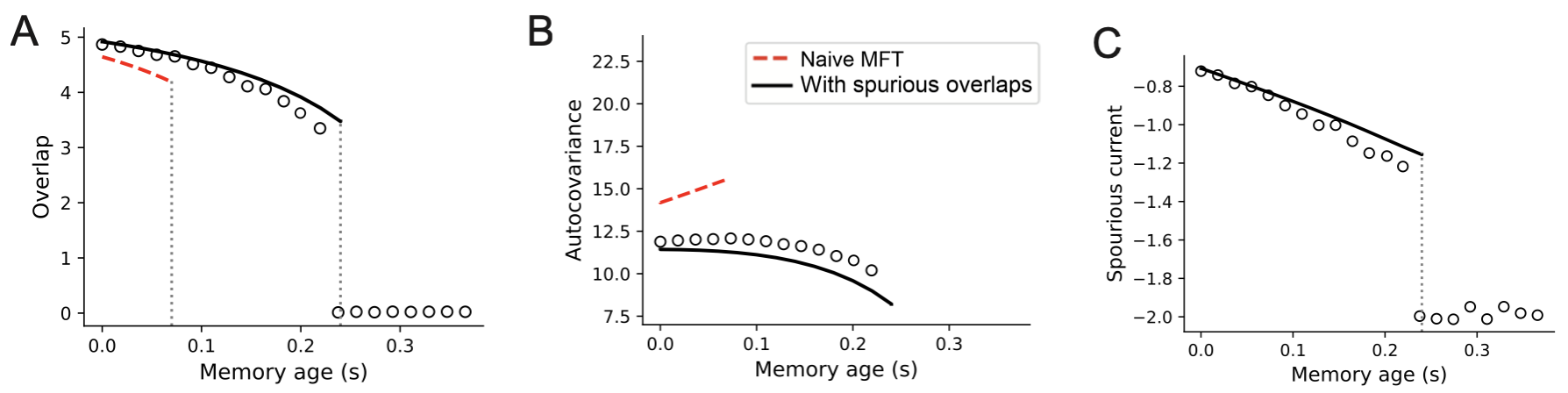}
\caption{\small Comparison between DMFT results (full lines), numerical results from simulations (empty dots) and naive MFT (dashed lines), where the effect of the spurious overlaps is not considered. In all panels, the x axis represents $\mu/Nc$. Panel A shows the overlap $m^\mu$ during retrieval, panel B shows the autocovariance of the local fields $\Delta_0$ (which is equal to $\Delta_\infty$ as recall corresponds to a fixed point attractor for the selected values of the parameters), panel C shows the corresponding spurious current term $S$. DMFT predicts a static retrieval phase until a pattern age $t\sim 0.23$, followed by non-retrieval. 
The size of the simulated network is $N=3\cdot10^5$; other parameters are $a=1.7$, $b=3.55$.} 
\label{fig:m_VS_t_static}
\end{figure*}

We now turn to a description of how a non-zero value of $S$ affects storage capacity. In \cref{fig:m_VS_t_static},the results of the DMFT that takes into account this contribution of spurious overlaps to mean inputs are compared with simulations, and with an incorrect \emph{naive DMFT}, where the role of spurious overlaps is neglected by setting $S=0$ in the Mean Field self-consistent equations. In \cref{fig:m_VS_t_static}, we choose parameters (see caption) for which the network converge to fixed point attractors. 
In \cref{sec:Solution_DMFT_stability}, we show similar plots, with parameters for which the network exhibits chaotic attractors. In general, increasing the value of the learning rate $b$ leads to a larger chaotic region.
Simulations and DMFT are in good agreement, both with regard to the quality of retrieval, and the fixed-point or chaotic nature of the corresponding attractor. Panel A displays the overlap $m^\mu$ as a function of $\mu/Nc$. The DMFT curve, as the simulations results, undergo a first order transition to $m^\mu=0$ for sufficiently large $\mu$. Similar discontinues transitions can be seen in the other order parameters $\Delta$ (panel B) and $S$ (panel C).

\Cref{fig:m_VS_t_static} shows that the performance of the network is strongly degraded in the naive DMFT, both in terms of capacity and quality of retrieval. This effect can be understood intuitively using the following reasoning. For each pattern, we partition the neuron population into two classes: \emph{foreground neurons}, defined as those whose firing rates during learning exceed the inflection point of \( f(\cdot) \), and \emph{background neurons}, whose firing rates remain below this point. Now, imagine starting from a solution of the DMFT, and then setting $S=0$, while keeping all other order parameters fixed. Since $S$ shifts the local fields to lower values, all local fields will increase. This has a weak impact on the firing rates of foreground neurons, as during recall they experience local fields that are stronger than the threshold of the activation function $\Phi$. On the other hand, background neurons experience local fields that are typically lower than the threshold of $\Phi(\cdot)$. Increasing these local fields by setting $S=0$ puts them closer to or even above threshold, thereby significantly increasing their firing rates. In turn, this higher network activity increases the variance of the noise field (see \cref{sec:Static_Mean_Field_Theory} for more details).

\begin{figure}[ht!]
\includegraphics[width=0.8\linewidth]{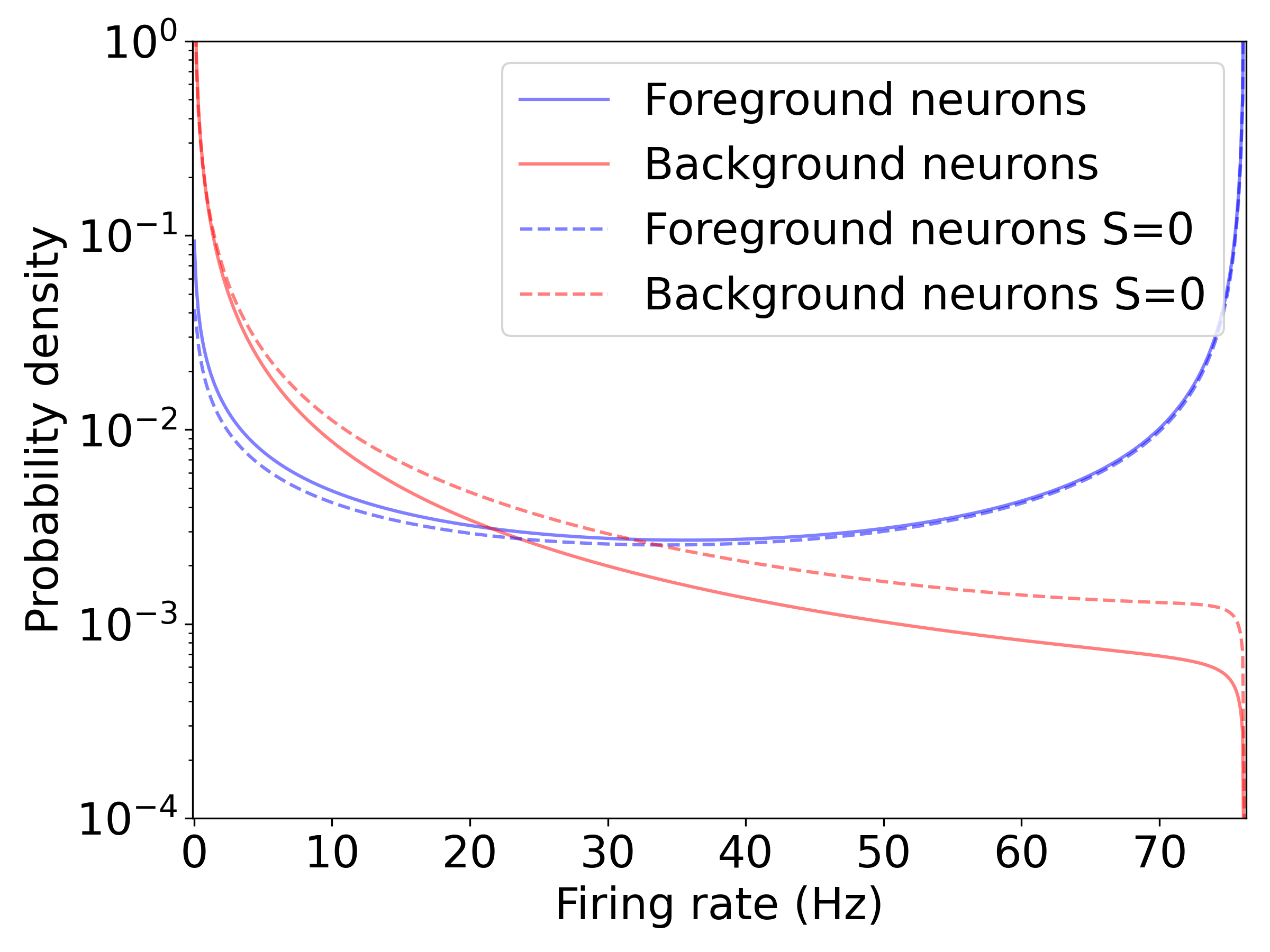}
\caption{\small Probability distribution of firing rates during recall for foreground (blue) and background (red) neurons. Full lines refer to the complete DMFT, while dotted colored lines are obtained by setting $S=0$.}
\label{fig:2}
\end{figure}
The effect of $S$ on the distribution of firing rates of background (red) and foreground (blue) neurons is depicted in \cref{fig:2}.  
The figure shows that including the effect of $S$ significantly reduces the chance of a background neuron exceeding the threshold, while the behavior of foreground neurons is largely unchanged. While for the complete DMFT background neurons account for only $3-5\%$ of neurons that exceed half of the maximum firing rate, this fraction increases to $20-25\%$ when $S$ is set to zero. Notice that, although not evident in the figure, probability density goes to zero at the maximum firing rate, for all curves. 

Notice that the importance of spurious correlations during retrieval has gone unnoticed in the previous literature on Hopfield-like neural networks, since the post-synaptic activation function $f(\cdot)$ was taken to have zero mean, leading to $S=0$. In our model, in which we used a function $f(\cdot)$ inferred from neurophysiological data, $S$ is instead negative, leading to a large increase in storage capacity.

\section{Phase diagram of the model}
\label{sec:Phase_diagram}
\begin{figure}
        \includegraphics[width=0.8\linewidth]{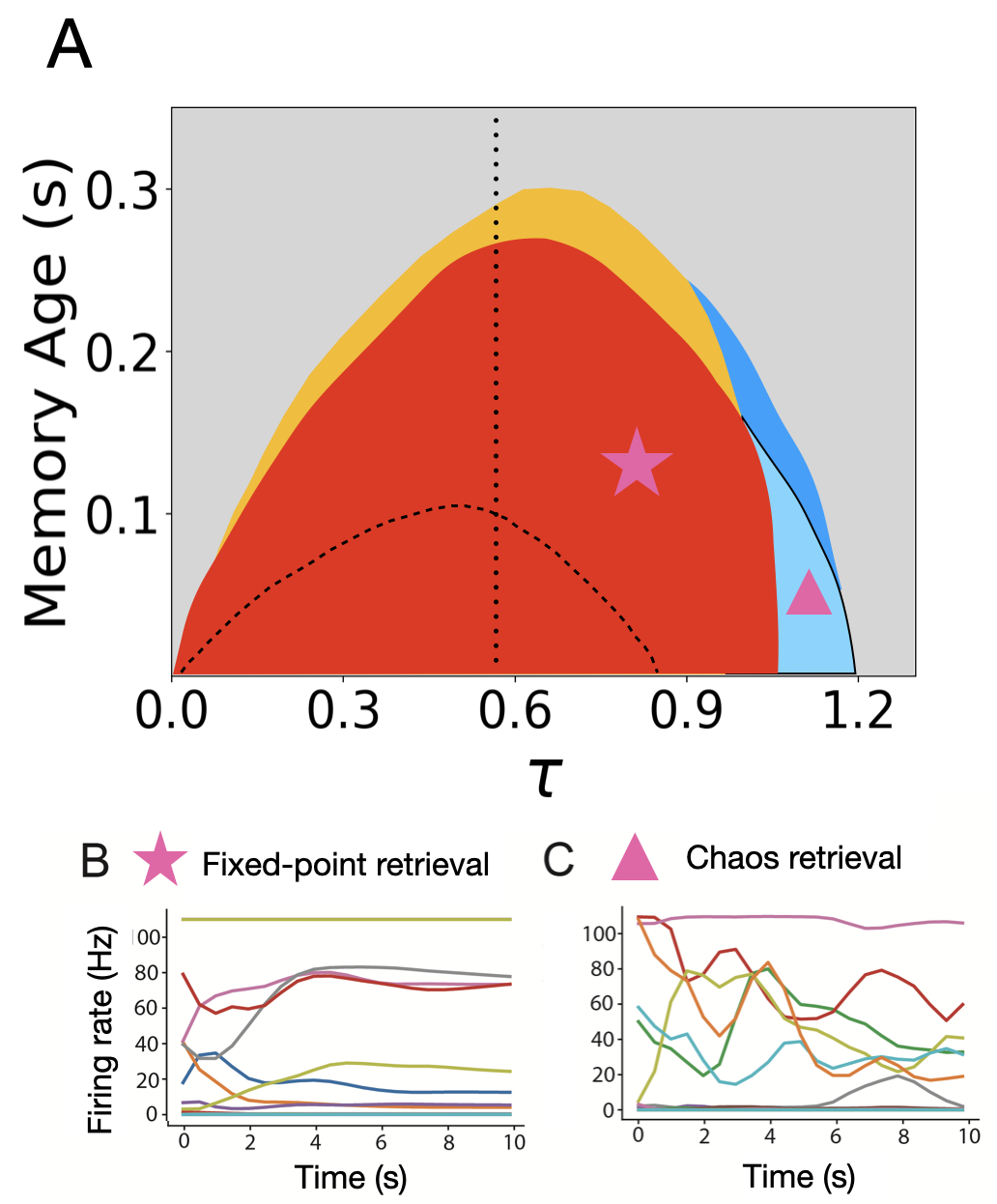}
    \caption{\small A) Phase diagram of the model, for $b=3.55$. Red: static retrieval, stable; Yellow: static retrieval, unstable; Light Blue: chaotic retrieval, stable; Blue: chaotic retrieval, unstable; Grey: no retrieval solution. The dashed line corresponds to the capacity predicted by the naive MFT, where $S=0$. The dotted line corresponds to the parameters in \cref{fig:m_VS_t_static}. 
        B) 20 simulated neuron trajectories, in the static recall phase. C) 20 simulated neuron trajectories, in the chaotic recall phase.
        }
        \label{fig:phase_diagram}
\end{figure}

We now explore the different phases of the model as a function of age $\mu/(cN)$ and forgetting time constant $\tau$. \cref{fig:phase_diagram} shows the boundaries of the different regimes, obtained from DMFT. For very fastforgetting timescales, the network is able to recall only a small number of stored memories, and the capacity goes to zero in the $\tau\rightarrow 0$ limit. In the opposite slow forgetting limit, the connectivity matrix stores too many memories at the same time, leading to a `blackout catastrophe' as in the Hopfield model and related models.

For intermediate values of $\tau$, the network is able to retrieve an extensive number of the most recently stored memories. At low values of memory age $s$ and synaptic decay time constant $\tau$, the network dynamics settles to fixed point attractors that are strongly correlated with stored patterns, characterized by $\Delta_0=\Delta_\infty$ (red area in \cref{fig:phase_diagram}). Increasing $\tau$ leads to a transition to chaotic attractors states in which rates never settle to fixed values, maintaining a finite correlation with the stored memories (light blue region in \cref{fig:phase_diagram}). For any value of $\tau$ there exist a specific age $t$ beyond which the network is unable to retrieve older memories and instead converges to the most recently stored memory, regardless of the initialization. We refer to this time as ``capacity $t_c$''. This transition is marked by the onset of instability of the $m\neq0$ solution of the DMFT w.r.t. perturbations towards the most recently stored pattern (see \cref{sec:Solution_DMFT_stability}). In \cref{fig:phase_diagram} static and chaotic unstable solutions are marked in yellow and blue respectively. Panels B and C display representative examples of the evolution over time of $20$ randomly chosen neurons, obtained in numerical simulations of the network. Panel B shows an example of static retrieval, where rates settle to a fixed value. Panel C shows instead an example of chaotic retrieval, with pronounced rate fluctuations. In both cases, only a minority of neurons display high firing rates, consistent with a sparse encoding.

Interestingly, as noted in ref.~\cite{pereira2018attractor} in the context of a non-forgetful network, the learning rule inferred from neurophysiological data is close to optimal w.r.t. the choice of the parameters of $f(\cdot),\,g(\cdot)$. (see Appendix\cref{sec:Optimal_learning_rules} for details). The performance of the network degrades sharply as one moves away from the optimal parameters. An exploration of parameter space reveals how model performance is quickly degraded when $\langle f(\tilde{r}) \rangle$ gets close or above zero (i.e. above the red line in \cref{fig:parameters_optimality_a_b}). Since $S\sim \langle f(\tilde{r}) \rangle$, this degradation corresponds to reducing the beneficial feedback from the spurious current, or even reversing it to $S>0$.

\section{Conclusions}
In this work, we developed a Dynamical Mean Field Theory (DMFT) framework to analyze the role of spurious overlaps in attractor neural networks governed by biologically inspired learning rules. Contrary to classical models, where spurious overlaps contribute to the variance but not the mean of synaptic inputs to neurons, we demonstrated that in these more realistic networks, the collective effect of spurious overlaps suppresses the mean input current, promoting sparse neural activity and improving memory capacity.

Our analysis revealed that this mechanism is driven by learning rules, in which the POST function describing the dependence of the rule on post-synaptic firing rate has a negative mean over the distribution of stored patterns. This contrasts with classic covariance-type learning rules, such as the one used in the Hopfield model, where spurious overlaps have no net effect on mean synaptic inputs. By leveraging the cumulative contribution of spurious overlaps, networks with learning rules with negative average post-synaptic terms achieve higher storage capacity and better retrieval performance.

We also explored the stability of solutions to the DMFT equations, identifying regimes of static and chaotic retrieval. Static solutions correspond to fixed-point attractors, while chaotic solutions are characterized by persistent fluctuations in neural activity. Our results show that chaotic solutions are often the stable ones, particularly for older patterns, and play a critical role in extending the network's memory capacity. 

The role of the mean spurious current $S$ emerged as a key factor in regulating network dynamics. By suppressing the activity of background neurons, $S$ reduces the overall noise in the system, enabling the network to maintain robust retrieval even in the presence of interference from other stored patterns. This self-regulating mechanism highlights the importance of spurious overlaps as a homeostatic feature of the network, rather than purely a source of degradation.

Finally, we showed that the learning rules inferred from biological data are near-optimal for memory storage. These rules exploit the beneficial effects of spurious overlaps, achieving a balance between sparsity and capacity. Our findings suggest that the design principles of biological networks may have evolved to harness the collective effects of spurious overlaps, providing a new perspective on the mechanisms underlying associative memory.
\bibliography{PRL_draft}


\clearpage
\onecolumngrid
\appendix
\section{Static Mean Field Theory}
\label{sec:Static_Mean_Field_Theory}
Let us recall the general setting of the Static Mean Field approach. During the training phase, the network learns according to \cref{eq:coupling_matrix} rate configurations $\tilde{r}_i^\mu=\Phi(\xi^\mu_i)$. During recall, the network is initialized to a state strongly correlated to one of the memories, and then let to relax to a fixed point of the dynamics:
\begin{equation}
    \begin{cases}
        \dot{r}_i=-r_i+\Phi\left(\sum_j c_{ij}J_{ij}r_j\right)\\
        r_i(t=0)=\tilde{r}_i^\mu
    \end{cases}
    \xRightarrow{\lambda_i(t):=\sum_j c_{ij}J_{ij}r_j(t)} \quad
    \begin{cases}
        \dot{\lambda}_i=-\lambda_i+\sum_j c_{ij}J_{ij}\Phi(\lambda_j)\\
        \lambda_i(t=0)=\sum_j c_{ij}J_{ij} \tilde{r}_j^\mu
    \end{cases}
\end{equation}
At a fixed point, rates obey
\begin{equation}
        r_i=\Phi\Big(\sum_{\alpha} e^{-\frac{a\alpha}{Nc}}\frac{b}{Nc}f(\tilde{r}^\alpha_i)\sum_{j\neq i}c_{ij} g(\tilde{r}^\alpha_j) r_j \Big) =: \Phi(\lambda_i).
\end{equation}
We are interested in the average value of the overlap order parameter
\begin{equation}
    m^\nu=\Big\langle\frac{1}{N}\sum_i g(\tilde{r}^\nu_i) r_i\Big\rangle,
\end{equation}
where $\langle \cdot \rangle$ represents an average with respect to the statistics of disorder (both structural connectivity matrix $c$ and stored patterns). It will be useful to set the following convention for pattern indexes:
\begin{itemize}
    \item $\mu$ refers exclusively to the pattern the dynamics was initialized close to.
    \item $\nu$ refers exclusively to the pattern we are computing the overlap with.
    \item $\alpha$ might refer to any pattern.
\end{itemize}
Out of the sum over patterns that builds the local field, two terms deserve special attention. One corresponds to pattern $\mu$, and is singled out by our initialization of the network; the other corresponds to pattern $\nu$ and is singled out by the fact that it is correlated to $\tilde{r}_\nu$;  (notice that this second term need not be considered if $\nu=\mu$):

\begin{equation}
    \begin{aligned}
        \lambda_i&= \sum_{\alpha} e^{-\frac{a\alpha}{Nc}}\frac{b}{Nc}f(\tilde{r}^\alpha_i)\sum_{j\neq i}c_{ij} g(\tilde{r}^\alpha_j) r_j\\
        &= \sum_{\alpha\neq \mu,\nu} e^{-\frac{a\alpha}{Nc}}\frac{b}{Nc}f(\tilde{r}^\alpha_i)\sum_{j\neq i}c_{ij} g(\tilde{r}^\alpha_j) r_j + e^{-\frac{a\nu}{Nc}}\frac{b}{Nc}f(\tilde{r}^\nu_i)\sum_{j\neq i}c_{ij} g(\tilde{r}^\nu_j) r_j + e^{-\frac{a\mu}{Nc}}\frac{b}{Nc}f(\tilde{r}^\mu_i)\sum_{j\neq i}c_{ij} g(\tilde{r}^\mu_j) r_j
    \end{aligned}
\end{equation}
For ease of exposition, let us add and subtract the average of each term in the sum over $\alpha$:
\begin{equation}
    \begin{aligned}
        \lambda_i= &\sum_{\alpha\neq \mu,\nu} \Big(e^{-\frac{a\alpha}{Nc}}\frac{b}{Nc}f(\tilde{r}^\alpha_i)\sum_{j\neq i}c_{ij} g(\tilde{r}^\alpha_j) r_j -e^{-\frac{a\alpha}{Nc}}b\langle f(\tilde{r})\rangle m^\alpha \Big) + 
        \sum_{\alpha\neq \mu,\nu} e^{-\frac{a\alpha}{Nc}}b\langle f(\tilde{r})\rangle m^\alpha +\\  
        &+ e^{-\frac{a\nu}{Nc}}\frac{b}{Nc}f(\tilde{r}^\nu_i)\sum_{j\neq i}c_{ij} g(\tilde{r}^\nu_j) r_j + e^{-\frac{a\mu}{Nc}}\frac{b}{Nc}f(\tilde{r}^\mu_i)\sum_{j\neq i}c_{ij} g(\tilde{r}^\mu_j) r_j
    \end{aligned}
\end{equation}
Now:
\begin{itemize}
    \item the first sum over $\alpha\neq \mu,\nu$ has zero average by construction, and converges to a Gaussian by the central limit theorem, since contributions coming from different sites are weakly correlated:
    \begin{equation}
        \sum_{\alpha\neq \mu,\nu} \Big(e^{-\frac{a\alpha}{Nc}}\frac{b}{Nc}f(\tilde{r}^\alpha_i)\sum_{j\neq i}c_{ij} g(\tilde{r}^\alpha_j) r_j -e^{-\frac{a\alpha}{Nc}}b\langle f(\tilde{r})\rangle m^\alpha \Big) =: h\sim \mathcal{N}(0, \Delta)
        \label{eq:noise_field}
    \end{equation}
    \item the second sum over $\alpha\neq \mu,\nu$ deserves a bit of thought. It is obviously zero if one assumes that spurious overlaps are zero $m^\alpha=0,\,\alpha\neq \mu$. If spurious overlaps were finite, on the other hand, it would be divergent, as it sums over an infinite stream of patterns. Finally, if spurious overlaps were $m^\alpha=O(1/Nc)$ for $\alpha\neq \mu$, the scaling would be just right to make it finite, since there are of order $(Nc)$ terms contributing to the sum, because of the exponential decay term. Let's call this term the \textit{spurious current} $S$:
    \begin{equation}
        \sum_{\alpha\neq \mu,\nu} e^{-\frac{a\alpha}{Nc}}b\langle f(\tilde{r})\rangle m^\alpha =: S
        \label{eq:SMFT_S}
    \end{equation}
    \item the $\nu$ pattern contribution to the local field constitutes a small correction to the local field. In this term, one might be tempted to trade $1/Nc\sum_{j\neq i}c_{ij} g(\tilde{r}^\nu_j) r_j$ for its average $m^\nu$, but that would be incorrect. In fact, we will see that $m^\nu\sim O(1/Nc)$, so that stochastic fluctuations cannot be neglected wrt the average:
    \begin{equation}
        e^{-\frac{a\nu}{Nc}}\frac{b}{Nc}f(\tilde{r}^\nu_i)\sum_{j\neq i}c_{ij} g(\tilde{r}^\nu_j) r_j =: \delta h
    \end{equation}
    \item in the $\mu-$term, the sum over $j$ is self averaging, and can be traded by its average introducing negligible errors:
    \begin{equation}
            e^{-\frac{a\mu}{Nc}}\frac{b}{Nc}f(\tilde{r}^\mu_i)\sum_{j\neq i}c_{ij} g(\tilde{r}^\mu_j) r_j \longrightarrow e^{-\frac{a\mu}{Nc}}bf(\tilde{r}^\mu_i)m^\mu
    \end{equation}
\end{itemize}
In summary, the local field can be written as
\begin{equation}
    \lambda_i=h_i+\delta h_i + S + e^{-\frac{a\mu}{Nc}}bf(\tilde{r}^\mu_i)m^\mu.
\end{equation}
and the expression for the overlaps becomes
\begin{equation}
    m^\nu=\int d\Vec{\xi}\,Dh\,P(\xi^0, ...)\,g\left((\Phi(\xi^\nu_i)\right)\Phi\left(h+\delta h + S + e^{-\frac{a\mu}{Nc}}bf(\tilde{r}^\mu_i)m^\mu\right) 
\end{equation}
When one is interested in computing the main overlap $\nu=\mu$, one obtains
\begin{equation}
    m^\mu=\Big\langle g\big[ \Phi(\xi^\mu)\big]\Phi\Big(h+S+e^{-\frac{a\mu}{Nc}}bf[\phi(\xi^\mu)]m^\mu)\Big) \Big\rangle.
\end{equation}
On the other hand, when computing any other overlap $\nu\neq\mu$, $\delta h$ becomes the only source of correlation between $g(\tilde{r}^\nu)$ and the local field, which cannot be neglected. To obtain a closed equation for the overlap, one must expand $\Phi(\lambda)$ to second order in $\delta h$. The result is as follows.
\begin{equation}
    m^\nu=\frac{1}{2}\frac{b^2}{Nc}e^{-2\frac{a\nu}{Nc}}\frac{\big\langle\Phi''(S+h+ e^{-\frac{a\mu}{Nc}}b f(\tilde{r}^\mu)m^\mu)\big\rangle \big\langle g(\tilde{r}^\nu) f^2(\tilde{r}^\nu) \big\rangle \big\langle g^2(\tilde{r}^\nu)r^2 \big\rangle}{1- b e^{-\frac{a\nu}{Nc} }  \big\langle \Phi'(S+h+ e^{-\frac{a\mu}{Nc}}b f(\tilde{r}^\mu)m^\mu) \big\rangle \big\langle g(\tilde{r}^\nu) f(\tilde{r}^\nu) \big\rangle}.\label{eq:spurious_overlaps1}
\end{equation}
The variance of the noise field $\Delta_0$ can be determined self-consistently from the definition \cref{eq:noise_field}, leading to the following set of self-consistent SMFT equations:  
\begin{align}
    &m^\mu=\Big\langle g\big[ \Phi(I^\mu)\big]\Phi\big(\lambda^\mu\big) \Big\rangle \\ 
    &m^\nu=\frac{1}{2}\frac{b^2}{Nc}e^{-2\frac{a\nu}{Nc}}\frac{\big\langle\Phi''(\lambda^\mu)\big\rangle \big\langle g(\tilde{r}^\nu) f^2(\tilde{r}^\nu) \big\rangle \big\langle g^2(\tilde{r}^\nu)\Phi^2(\lambda^\mu) \big\rangle}{1- b e^{-\frac{a\nu}{Nc} }  \big\langle \Phi'(\lambda^\mu) \big\rangle \big\langle g(\tilde{r}^\nu) f(\tilde{r}^\nu) \big\rangle} \\ 
    &S = \sum_{\alpha\neq \mu} e^{-\frac{a\alpha}{Nc}}b\langle f(\tilde{r})\rangle m^\alpha \label{eq:Sdef2}\\  
    &\Delta_0 = w\big\langle   \Phi^2(\lambda^\mu)\,\big\rangle
\end{align}
where $ w=\frac{b^2}{2a}\langle f^2(\tilde{r})\rangle \langle g^2(\tilde{r})\rangle$. Finally, substituting \cref{eq:spurious_overlaps1}, into \cref{eq:Sdef2} one gets
\begin{align}
    &m^\mu=\Big\langle g\big[ \Phi(I^\mu)\big]\Phi\big(\lambda^\mu\big) \Big\rangle \\ 
    &S = \frac{1}{2} b^3\langle f(\tilde{r}^\nu) \rangle \big\langle\Phi''(\lambda^\mu)\big\rangle \big\langle g(\tilde{r}) f^2(\tilde{r}) \big\rangle \big\langle g^2(\tilde{r})\Phi^2(\lambda^\mu) \big\rangle \int_0^{\infty} ds\frac{ e^{-3as} }{1- b e^{-as\nu}   \big\langle \Phi'(\lambda^\mu) \big\rangle \big\langle g(\tilde{r}) f(\tilde{r})\big\rangle } \label{eq:ave_s1} \\ 
    &\Delta_0 = w\big\langle   \Phi^2(\lambda^\mu)\,\big\rangle,
\end{align}

\Cref{eq:spurious_overlaps1} determines the average value of spurious overlaps $m^\nu$, which do not appear in any other equation. 
These spurious overlaps in turn influence other order parameters through $S$, which is determined self consistently from \cref{eq:ave_s1}. Even though spurious overlaps have large fluctuations compared to their average value $m^\nu$, an extensive number of them enters in $S$, so that knowing $m^\nu$ characterizes the system completely. 

\section{Dynamic Mean Field Theory}
\label{sec:Dynamic_Mean_Field_Theory}
Recall that local fields evolve according to  
\begin{equation}
        \dot{\lambda}_i=-\lambda_i+\sum_j c_{ij}J_{ij}\Phi(\lambda_j) \label{eq:field_dynamics}
\end{equation}
As discussed in \cite{sompolinsky1988chaos, kree87}, continuous rate models with asymmetric connectivity often admit solutions which are not static. These solutions can be captured through a DMFT approach (see e.g.~\cite{helias-dahmen-book}). The generating functional describing the system reads
\begin{equation}
  Z[s,\hat{s},c_{ij}, J_{ij}]=\int \mathcal{D}\lambda(t) \mathcal{D}\hat{\lambda}(t) \exp [-L_0-L],
\end{equation}
where 
\begin{equation}
\begin{aligned}
        &L_0=\int \sum_i \hat{\lambda}(\tau)[\lambda'_i(\tau)+\lambda_i(\tau)]d\tau-\sum_i\int [s_i(\tau)\hat{\lambda}_i(\tau)+\hat{s}_i(\tau)\lambda_i(\tau)] d\tau\Big]\\
        &L=-\int\sum_{ij}\hat{\lambda}_i(\tau)c_{ij}J_{ij}\Phi\big[\lambda_j(\tau)\big]d\tau.
\end{aligned}
\end{equation}
Averaging $L$ over $c_{ij}$ and expanding in powers of c one gets
\begin{equation}
  L=\prod_{ij} \Big[ 1-c\Big(1-\exp\big(\int\hat{\lambda}_i(\tau)J_{ij}\Phi\big[\lambda_j(\tau)\big]d\tau\big)\Big)\Big]=\exp\Big\{-\sum_{ij}c\Big[1-\exp\Big(\int \hat{\lambda}_i(\tau) J_{ij} \Phi\big[\lambda_j(\tau)\big]d\tau\Big)\Big] 
\end{equation}
Then, one expands the exponentials in powers of $J$ to second order, obtaining
\begin{equation*}
    \begin{aligned}
        L\sim \int d\tau &\sum_{ij} \hat{\lambda}_i(\tau) c J_{ij} \Phi\big[\lambda_j(\tau)\big]+ 
        \int d\tau d\tau' \sum_{ij}\hat{\lambda}_i(\tau) \Big[\frac{c}{2}\sum_{hk}J_{ih}J_{jk}\Phi\big[\lambda_h(\tau)\big] \Phi\big[\lambda_k(\tau')\big]\delta_{ij}\delta_{hk}\big]\Big] \hat{\lambda}_j(\tau') \\
        = \int d\tau &\sum_i \hat{\lambda}_i(\tau) \sum_\mu b e^{-\frac{a\mu}{Nc}}f\big(\tilde{r}^\mu_i\big)m^\mu + \frac{1}{2}\int d\tau d\tau'\sum_{ij}\hat{\lambda}_i(\tau) \Big[C_{ij}(\tau-\tau')\Big] \hat{\lambda}_j(\tau')
    \end{aligned}
\end{equation*}
where we have used \cref{eq:coupling_matrix} for the couplings and defined $C_{ij}\,$ and the overlaps $m^\alpha(\tau)$ as
\begin{align}
    m^\alpha(\tau)=\frac{1}{N}&\sum_j g\big(\tilde{r}^\alpha_j\big)\Phi\big[\lambda_j(\tau)\big] \label{eq:all_m}\\
    C_{ij}(\tau-\tau')&= c\sum_{hk}J_{ih}J_{jk}\Phi\big[\lambda_h(\tau)\big]\Phi\big[\lambda_k(\tau')\big] \nonumber \\
    &=c\sum_{\alpha\beta}\frac{b^2}{(Nc)^2}e^{-\frac{a(\alpha+\beta)}{Nc}} f\big(\tilde{r}^\alpha_i\big) f\big(\tilde{r}^\alpha_j\big) \sum_{hk} \Phi\big[\lambda_h(\tau)\big]\Phi\big[\lambda_k(\tau')\big] g\big(\tilde{r}^\alpha_h\big) g\big(\tilde{r}^\beta_k\big)\\
    &=c\sum_{\alpha}\frac{b^2}{(Nc)^2}e^{-\frac{2a\alpha}{Nc}} f\big(\tilde{r}^\alpha_i\big) f\big(\tilde{r}^\alpha_j\big)  \langle \Phi\big[\lambda_h(\tau)\big]\Phi\big[\lambda_h(\tau')\big]\rangle \langle g^2\big(\tilde{r}\big)\rangle\label{eq:C}
\end{align}
This action can be reinterpreted as describing a set of $N$ self-consistent stochastic differential equations
\begin{equation}
\label{eq:u_dynamics}
        \partial_\tau \lambda_i(\tau)=-\lambda_i(\tau)+\sum_\mu b e^{-\frac{a\alpha}{Nc}}f\big(\tilde{r}^\alpha_i\big)m^\alpha+\zeta_i(\tau)
\end{equation}
where $\zeta$ is a Gaussian random field, correlated according to
\begin{equation}
    \mathop{\mathbb{E}}[\zeta_i(\tau) \zeta_j(\tau')] = C_{ij}(\tau-\tau').
\end{equation}
The values of $m^\mu$ and $C_{ij}$ must be determined self-consistently, through \cref{eq:all_m,eq:C}. Introducing noise fields $h_i(\tau):= \lambda_i(\tau)-\sum_\mu b e^{-\frac{a\mu}{Nc}}f\big(\tilde{r}^\mu_i\big)m^\mu$, we obtain 
\begin{equation}
\label{eq:DMFT_not_split}
\addtolength\jot{6pt}
    \begin{aligned}
        &\partial_\tau h_i(\tau)=-h_i(\tau)+\zeta_i(\tau)\\
        &\mathop{\mathbb{E}}[\zeta_i(\tau) \zeta_j(\tau')] = C_{ij}(\tau-\tau')\\
        & m^\alpha(\tau)=\frac{1}{N}\sum_j g\big(\tilde{r}^\alpha_j\big)\Phi\big[h_j(\tau)+\sum_\beta b e^{-\frac{a\beta}{Nc}}f\big(\tilde{r}^\beta_j\big)m^\beta\big], \qquad \alpha=1,...,P.\\
     \end{aligned}
\end{equation}   
The expression for $C_{ij}(\tau-\tau')$ is self-averaging, 
and fluctuations of $C_{ij}$ around its average value shrink to zero in the limit $N\to\infty$. If one looks for solutions of the form $m=(0,...,m^\mu,0,...)$, fluctuations of $C_{ij}$ can be neglected, and $C_{ij}$ can be replaced by its averaged value, obtaining a system of three self-consistent equations: 
\begin{equation}
\addtolength\jot{6pt}
\label{eq:DMFT_leading_only}
    \begin{aligned}
        &\partial_\tau h_i(\tau)=-h_i(\tau)+\zeta_i(\tau)\\
        &\mathop{\mathbb{E}}[\zeta_i(\tau) \zeta_j(\tau')] = w\big\langle \Phi\big[h(\tau)+ b e^{-\frac{a\mu}{Nc}}f\big(\tilde{r}^\mu_h\big)m^\mu\big]\Phi\big[h(\tau')+ b e^{-\frac{a\mu}{Nc}}f\big(\tilde{r}^\mu_k\big)m^\mu\big]\big\rangle \delta_{ij}\\
        & m^\mu=\big\langle g\big(\tilde{r}^\mu_i\big)\Phi\big[h+ b e^{-\frac{a\mu}{Nc}}f\big(\tilde{r}^\mu_i\big)m^\mu\big]\big\rangle
    \end{aligned}
\end{equation}
where we have assumed that, at equilibrium, $m^\mu(\tau)$ is independent on $\tau$. The noise fields $h_i(\tau)$ are then i.i.d. Gaussian variables, whose asymptotic statistics can be characterized by the autocorrelation $\Delta(t):=\lim_{s\to\infty}\langle h_i(s)h_i(s+t) \rangle$. Differentiation with respect to $t$ in light of \cref{eq:u_dynamics} proves that $\Delta(t)$ solves a second order differential equation, analogous to a one dimensional motion in a potential \cite{kree87, tirozzi1991chaos,crisanti18}:
\begin{equation}
        \Delta''(t)=\langle h'(s)h'(s+t) \rangle =\langle u'(s)u'(s+t) \rangle=\Delta(t)-\mathop{\mathbb{E}}[\zeta_i(\tau) \zeta_i(\tau')]=-\partial_\Delta V(\Delta, \Delta_0)
\end{equation}
with initial condition $\Delta(0)=\Delta_0\,$,  $\Delta'(0)=0$. The explicit expression of the potential is
\begin{equation}
\label{eq:potential}
\begin{aligned}
        V(\Delta, \Delta_0)=&-\frac{\Delta^2}{2}+w\big\langle \tilde{\Phi}\big[h_0+ b e^{-\frac{a\mu}{Nc}}f\big(\tilde{r}^\mu\big)m^\mu\big]\tilde{\Phi}\big[h_{t}+ b e^{-\frac{a\mu}{Nc}}f\big(\tilde{r}^\mu\big)m^\mu\big]\big\rangle
        \end{aligned}
\end{equation}
where $\tilde{\Phi}$ is a primitive of the transfer function $\partial_x\tilde{\Phi}(x)=\Phi(x)$, and $h_0$, $h_{t}$ are Gaussian fields, with variance ${\langle h^2_0\rangle=\langle h^2_t\rangle=\Delta_0}$ and covariance $\langle h_0 h_{t}\rangle=\Delta(t)$. Hence, the parameter $\Delta_0$ fixes both the initial position of the “dynamics” and the shape of the potential through \cref{eq:potential}. Different choices of $\Delta_0$ parametrize different solutions to the DMFT. The only stable solutions correspond to an asymptotic value of the autocorrelation function $\Delta_\infty:=\lim_{t\to\infty}\Delta(t)$ which is a fixed point of the dynamics (see \cite{kree87, sompolinsky1988chaos, crisanti18}). In formulae
\begin{equation}
    \begin{aligned}
    \begin{cases}
        &\partial_{\Delta}V(\Delta_\infty, \Delta_0)=0 \\
        &V(\Delta_\infty,\Delta_0)=V(\Delta_0,\Delta_0)\\
    \end{cases}
    \end{aligned}
\end{equation}
which can be closed by the self-consistency equation for the only non-zero $m$, leading to
\begin{equation}
    \begin{aligned}
    \addtolength\jot{6pt}
    \begin{cases}    
        &\Delta_\infty=w\big\langle {\Phi}\big[h_0+ b e^{-\frac{a\mu}{Nc}}f\big(\tilde{r}^\mu\big)m^\mu\big]{\Phi}\big[h_{\infty}+ b e^{-\frac{a\mu}{Nc}}f\big(\tilde{r}^\mu\big)m^\mu\big]\big\rangle\\
        &\Delta_\infty^2-2w\big\langle \tilde{\Phi}\big[h_0+ b e^{-\frac{a\mu}{Nc}}f\big(\tilde{r}^\mu\big)m^\mu\big]\tilde{\Phi}\big[h_{\infty}+ b e^{-\frac{a\mu}{Nc}}f\big(\tilde{r}^\mu\big)m^\mu\big]\big\rangle=\\
        &\Delta_0^2-2w\big\langle \tilde{\Phi}^2\big[h_0+ b e^{-\frac{a\mu}{Nc}}f\big(\tilde{r}^\mu\big)m^\mu\big]\big\rangle \\
        &m^\mu=\big\langle g\big(\tilde{r}^\mu\big)\Phi\big[h_0+ b e^{-\frac{a\mu}{Nc}}f\big(\tilde{r}^\mu\big)m^\mu\big]\big\rangle
        \end{cases}
    \end{aligned}
\end{equation}
where $h_0$, $h_{\infty}$ are Gaussian fields, with variance ${\langle h^2_0\rangle=\langle h^2_{\infty}\rangle=\Delta_0}$ and covariance $\langle h_0 h_{\infty}\rangle=\Delta_\infty$. \\
In our case, we are interested in solutions where component $\mu$ of $m$ is finite in the thermodynamic limit (the one corresponding to the pattern we have initialized the net closest to), and others are $O(1/Nc)$. Such small contributions to the formulation are, as in the SMFT case, represented cumulatively by the inclusion of a \textit{shift} in the local field
\begin{equation}
    S:=\sum_{\alpha\neq\mu} e^{-\frac{a\alpha}{Nc}} b f(\tilde{r}^\alpha_i)m^\alpha
\end{equation}
The equations describing the small spurious overlaps are sensitive to the subleading corrections to the local fields' statistics, due to fluctuations of \cref{eq:C} around its disorder averaged value. 
Specifically, one can split the stochastic force in \cref{eq:DMFT_not_split} as $\zeta_i(\tau)=\zeta^0_i(\tau)+\zeta^1_i(\tau)$, where
\begin{equation}
    \begin{aligned}
        \mathop{\mathbb{E}}[\zeta^0_i(\tau) \zeta^0_j(\tau')] &= w\big\langle \Phi\big[h(\tau)+ S+b e^{-\frac{a\mu}{Nc}}f\big(\tilde{r}^\mu_h\big)m^\mu\big]\Phi\big[h(\tau')+S+ b e^{-\frac{a\mu}{Nc}}f\big(\tilde{r}^\mu_k\big)m^\mu\big]\big\rangle \delta_{ij}\\
        \mathop{\mathbb{E}}[\zeta^1_i(\tau) \zeta^1_j(\tau')] &= C_{ij}(\tau-\tau')-\mathop{\mathbb{E}}[\zeta^0_i(\tau) \zeta^0_j(\tau')]
    \end{aligned}
\end{equation}
and do the same for the noise field $h=h^0+\delta h$, where
\begin{equation}
    \label{eq:DMFT_split}
    \begin{aligned}
        &m^\alpha=\big\langle g\big(\tilde{r}^\alpha_j\big)\Phi\big[h^0_j+\delta h_j+ S+  b e^{-\frac{a\mu}{Nc}}f\big(\tilde{r}^\mu_i\big)m^\mu\big]\big\rangle\\
        &\partial_\tau h^0_i(\tau)=-h^0_i(\tau)+\zeta^0_i(\tau) \\
        &\partial_\tau \delta h_i(\tau)=-\delta h_i(\tau)+\zeta^1_i(\tau) \\
        &\mathop{\mathbb{E}}[\zeta^0_i(\tau) \zeta^0_j(\tau')] = w\big\langle \Phi\big[h(\tau)+ S+ b e^{-\frac{a\mu}{Nc}}f\big(\tilde{r}^\mu_h\big)m^\mu\big]\Phi\big[h(\tau')+ S+ b e^{-\frac{a\mu}{Nc}}f\big(\tilde{r}^\mu_k\big)m^\mu\big]\big\rangle \delta_{ij}\\
        &\mathop{\mathbb{E}}[\zeta^1_i(\tau) \zeta^1_j(\tau')] = C_{ij}(\tau-\tau')-\mathop{\mathbb{E}}[\zeta^0_i(\tau) \zeta^0_j(\tau')]
    \end{aligned}
\end{equation}
Assuming that the correction $\delta h$ is small, the equations for the spurious overlaps can be expanded to second order in $\delta h$, leading to
\begin{equation}
    m^\alpha=\frac{1}{2}\frac{b^2}{Nc}e^{-\frac{2a\alpha}{Nc}}\frac{\big\langle\Phi''(\lambda_0)\big\rangle \big\langle g(\tilde{r}) f^2(\tilde{r}) \big\rangle \big\langle g^2(\tilde{r})\big\rangle \big\langle\Phi^2\big(\lambda_0\big)\big\rangle}{1- b e^{-\frac{a\alpha}{Nc} }  \big\langle \Phi'(\lambda_0) \big\rangle \big\langle g(\tilde{r}^\alpha) f(\tilde{r}^\alpha)\big\rangle},\quad \alpha\neq\mu \label{eq:spurious_overlaps}
\end{equation}    
where we have introduced the notation $\lambda_{0(\infty)}:=S+h_{0(\infty)}+e^{-\frac{a\mu}{Nc}}bf\big(\tilde{r}^\mu\big)m^\mu$. Substituting \eqref{eq:spurious_overlaps} in the equation for $S$, and taking the large $cN$ limit, one obtains the following set of DMFT equations 
\begin{align}
    &m^\mu=\Big\langle g\big(\tilde{r}^\mu\big)\Phi\big(\lambda_0\big) \Big\rangle\\
    &S = \frac{1}{2}b^3\langle f(\tilde{r}^\nu) \rangle\big\langle\Phi''(\lambda_0)\big\rangle \big\langle g(\tilde{r}) f^2(\tilde{r}) \big\rangle \big\langle g^2(\tilde{r})\big\rangle\big\langle\Phi^2\big(\lambda_0\big)\big\rangle \int_0^{\infty} ds\frac{e^{-3as}}{1- b e^{-as}\big\langle \Phi'(\lambda_0) \big\rangle \big\langle g(\tilde{r}) f(\tilde{r})\big\rangle }
    \label{eq:ave_s}\\
    &\Delta_\infty-2w\big\langle \widetilde{\Phi}\big(\lambda_0\big) \widetilde{\Phi}\big(\lambda_{\infty}\big) \big\rangle = \Delta_0-2w\big\langle \widetilde{\Phi}^2\big(\lambda_0\big)\big\rangle
    \label{eq:dmft4_bis}\\
    &\Delta_\infty = \big\langle   \Phi\big(\lambda_0\big) \,\Phi\big(\lambda_{\infty}\big)\big\rangle 
    \label{eq:dmft3} 
\end{align}
Now \cref{eq:spurious_overlaps} determines the average value of spurious overlaps $m^\nu$, which do not appear in any other equation. Their influence is summarized in the term $S$, which is determined self consistently from \cref{eq:ave_s}. The DMFT might admit \textit{static solutions}, i.e. solutions where $\Delta_\infty=\Delta_0$. In that case, equations reduce to those discussed in \cref{sec:Static_Mean_Field_Theory}. \textit{Chaotic solutions} are marked by $\Delta_\infty<\Delta_0$. Proving this rigorously will require computing the corresponding maximum Lyapunov exponent \cite{kadmon2015transition, crisanti18}.

\section{Solution of DMFT equations, and their stability}
\label{sec:Solution_DMFT_stability}

Once we find numerically all solutions to the DMFT, we are left with determining which ones are stable from a dynamical point of view. This can be done, for example, by considering the joint statistics of a \textit{replicated} system, comprising two independent copies of the network, evolving according to the same realization of the couplings (see \cite{sompolinsky1988chaos} for an example of this approach). From a practical point of view, one can obtain an accurate prediction for the stability of a solution through a simpler heuristic reasoning \cite{pereira2023forgetting}. The key point is noticing that, when the network is asked to retrieve an old pattern, it will sometimes converge to a fixed-point attractor correlated with one of the more recent memories instead. So, one can simplify the search for stable solutions by restricting the analysis to one kind of instability only, namely a small perturbation dragging the system towards the most recently learned pattern, in the form of a small (but finite in the thermodynamic limit) magnetization $m^0$. One should in principle develop the full dynamical theory describing the evolution over time $\tau$ of $m^\mu(\tau), m^0(\tau)$ and the other quantities. Fixed points of these dynamic equations correspond to solutions of the DMFT. As in \cite{pereira2023forgetting}, it turns out that assuming that local fields and $S$ relax instantly to their equilibrium values given $m(\tau)$ is a good approximation of this dynamics. In this setting, $(m^\mu(\tau), m^0(\tau))$ are the only dynamical variables, while all other quantities are
functions of $(m^\mu(\tau), m^0(\tau))$, determined by the DMFT equations \cref{eq:ave_s,eq:dmft3,eq:dmft4_bis} at each time-step. The equations of motion for the magnetizations can be derived by differenciating with respect to time the definition of $m$ \cref{eq:all_m}, and using the rates' dynamics \cref{eq:field_dynamics}, leading to
\begin{align}
            \label{eq:stab1}
            &\partial_{\tau}m^\mu(\tau)=-m^\mu(\tau)+ \Big\langle g\big[ \Phi(I^\mu)\big]\Phi\Big(h_0+S+e^{-at}bf[\phi(I^\mu)]m^\mu+bf[\phi(I^0)]m^0\Big) \Big\rangle \\
            &\partial_{\tau}m^0(\tau)=-m^0(\tau)+ \Big\langle g\big[ \Phi(I^0)\big]\Phi\Big(h_0+S+e^{-at}bf[\phi(I^\mu)]m^\mu+bf[\phi(I^0)]m^0\Big) \Big\rangle
\end{align}
The stability of this solution is determined by the hessian matrix of the RHS of the above equations, as functions of $m^\mu$ and $m^0$. The result of this analysis is that solutions generally become unstable as pattern age increases, before disappearing. The capacity of the network is very well predicted by the age of instability onset. On the other hand, the approximate stability criterion is not able to predict the instability of a static solution towards a chaotic solution. This distinction can be made only by analyzing the replicated system. Still, a review of similar models in the literature \cite{pereira2023forgetting, tirozzi1991chaos, kree87, gillett20} suggests that the chaotic solution is always the preferred one, whenever it exists. This is confirmed by simulations. \\

\begin{figure*}[ht!]
\includegraphics[width=\linewidth]{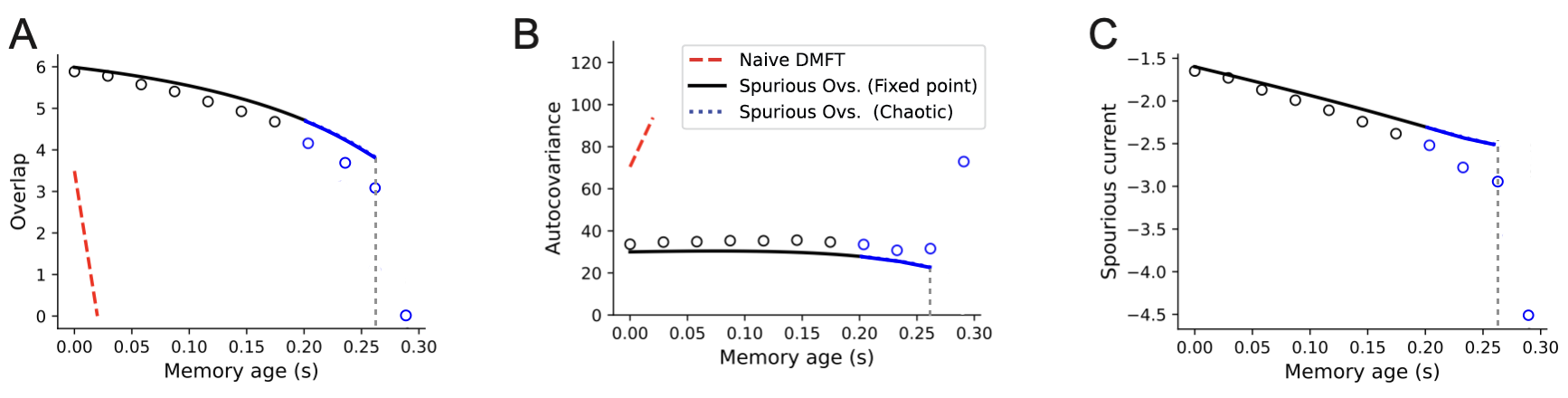}
\caption{As in \cref{fig:m_VS_t_static}, in a case where chaotic dynamics characterizes the retrieval of the oldest patters. Blue lines represent where DMFT predicts chaotic retrieval. Empty blue dots represent simulations that did not converge to a fixed point. The size of the simulated network is $N=3\cdot10^5$; other parameters are $a=1.3$, $b=4$.}
\label{fig:m_VS_t_chaos}
\end{figure*}

\section{Learning rules optimizing capacity}
\label{sec:Optimal_learning_rules}
\begin{figure}[t]        
\begin{subfigure}{0.48\linewidth}
\includegraphics[width=\linewidth]{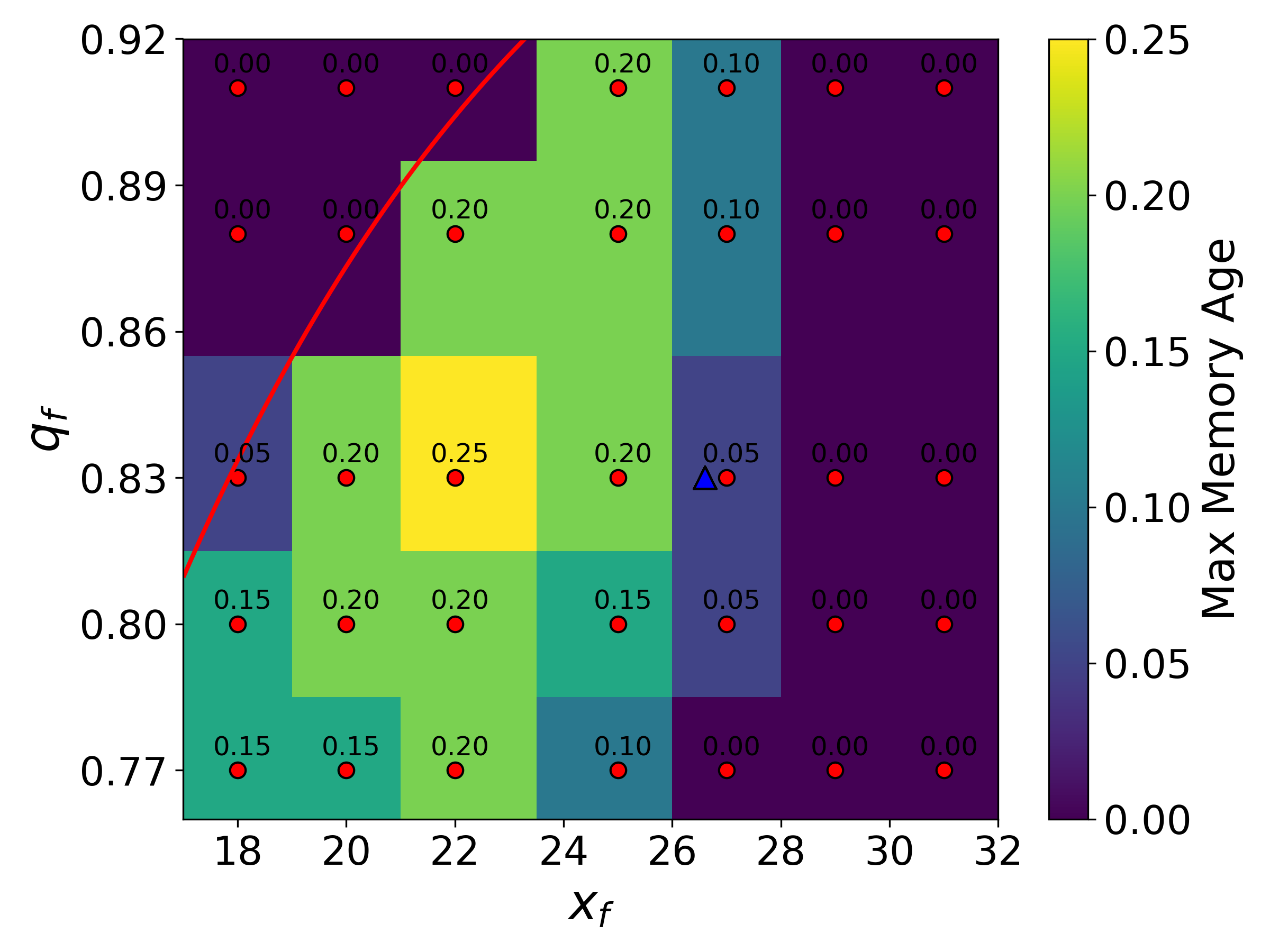}
\caption{}
\label{fig:parameters_optimality_xf_qf}
\end{subfigure}
\hfill
\begin{subfigure}{0.48\linewidth}
\includegraphics[width=\linewidth]{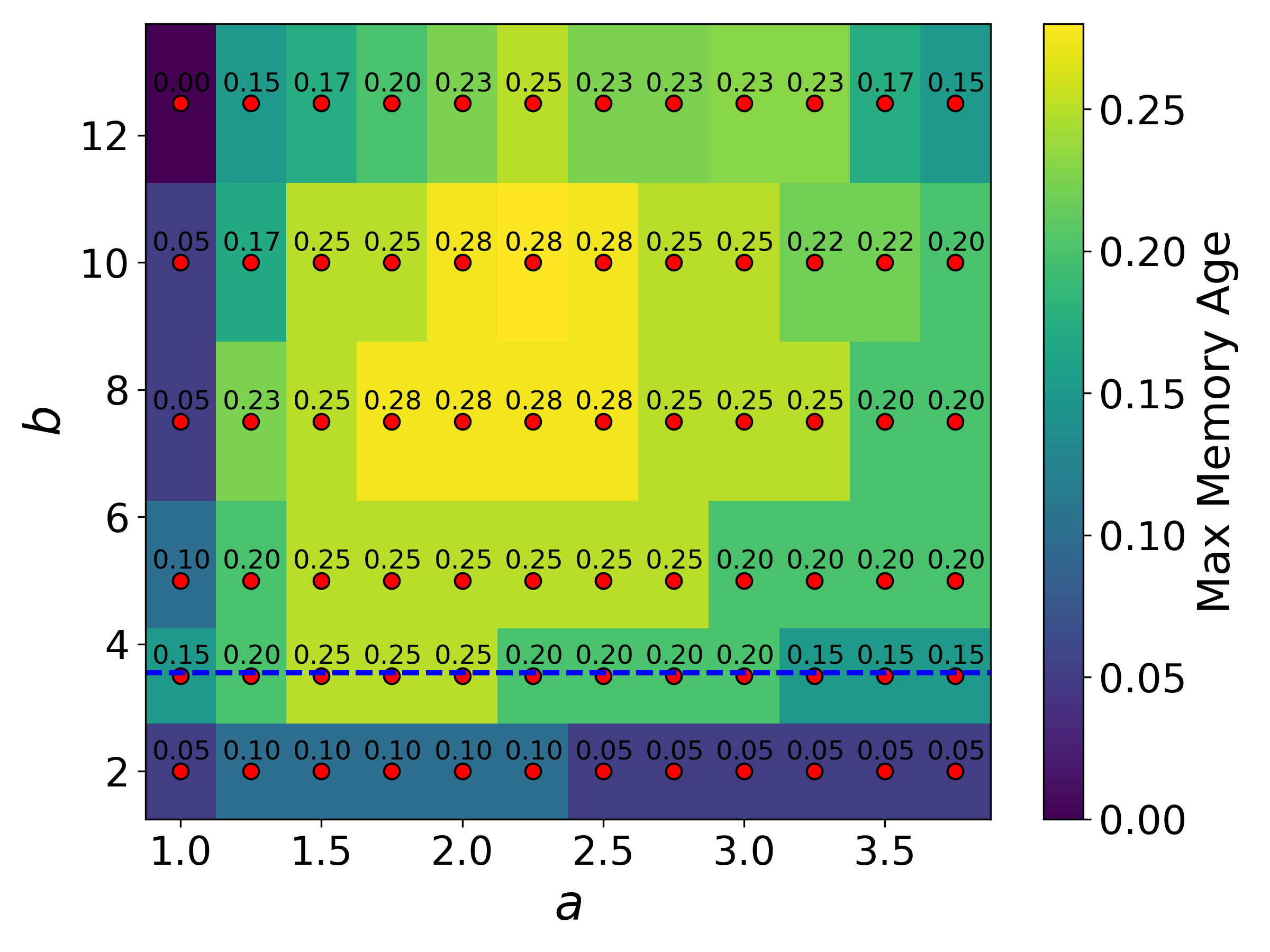}
\caption{}
\label{fig:parameters_optimality_a_b}
\end{subfigure}
\caption{Parameter optimization for network capacity. (a) Capacity of the network as a function of $q_f$ and $x_f$, for $a=1.3$ and $b=3.55$. Below the red line, $\langle f(\tilde{r})\rangle<0$. Capacity decreases dramatically above this line. The blue triangle indicates the value of the parameters inferred from in vivo data. (b) Capacity of the network as a function of $a$ and $b$, for $x_f=22$Hz and $q_f=0.83$. The blue dashed line corresponds  to the value of $b$ inferred from in vivo data. Interestingly, such biological value seems to be just as big as to give good capacity to the network.}
\label{fig:parameters_optimality}
\end{figure}

In ref.~\cite{pereira2018attractor}, it was noted that in the context of a non-forgetful network, the learning rule inferred from neurophysiological data was close to optimal w.r.t. the choice of the parameters of $f(\cdot),\,g(\cdot)$. In this section, we explore the parameter space of our model, varying $a,\,b$ and $x_f,q_f$ to verify whether this property holds true also when $f(\cdot),\,g(\cdot)$ are applied in the context of our palimpsestic learning rule.  \Cref{fig:parameters_optimality_xf_qf} shows there is qualitative agreement between the optimal parameters and the one inferred from biological data (blue triangle), with the biggest discrepancy being on $x_f$. Also, it is noteworthy that the performance of the network degrades sharply as one moves away from the optimal parameters. Exploring the $q_f,x_f$ space also reveals how model performance is quickly degraded if $\langle f(\tilde{r}) \rangle$ gets close or above zero (i.e. above the red line). Recall that $S\sim \langle f(\tilde{r}) \rangle$, hence this degradation corresponds to reducing the beneficial feedback from the spurious current, or even reversing it to $S>0$. \\
\Cref{fig:parameters_optimality_a_b} reports an analogous analysis in the space of learning rates and forgetting timescales. The value of learning rate inferred from data ($a=3.55$) seems to be just as big as to give good capacity to the network, but not as big as to give maximum capacity. This could be interpreted as an efficiency tradeoff, maximizing performance while keeping relatively low strength of synaptic connections, which come at an energy cost.

\end{document}